\documentclass[aps,prl,reprint,longbibliography]{revtex4-2}
\usepackage{amsmath}
\usepackage{amsfonts}
\usepackage{graphicx}
\usepackage{units}  
\usepackage{xspace}
\usepackage{subfigure}
\usepackage{hyperref}
\usepackage{enumitem}

\newcommand{\mb}{\mathbf}
\newcommand{\mk}{\mathbf{k}}
\newcommand{\mr}{\mathbf{r}}

\newcommand{\beq}{\begin{equation}}
\newcommand{\eeq}{\end{equation}}
\newcommand{\bea}{\begin{eqnarray}}
\newcommand{\eea}{\end{eqnarray}}
\newcommand{\mrm}{\mathrm}

\usepackage[usenames]{color}

\begin{document}

\title{Effect of nonlocal interlayer hopping on wave function in twisted bilayer graphene}

\author{Hridis K. Pal}
	\affiliation{Department of Physics, Indian Institute of Technology Bombay, Powai, Mumbai 400076, India}
\date{\today}

\begin{abstract}
The conventional low-energy theory employed to describe twisted bilayer graphene (TBG) relies on a local interlayer Hamiltonian. According to this theory, TBG has the same linear-in-momentum dispersion and spinor wave function at the Dirac point as single-layer graphene (SLG),  albeit with a renormalized velocity that decreases as the rotation angle between the layers decreases, eventually reaching zero at the magic angle. In this work, I expand upon this low-energy theory by including nonlocal terms in the interlayer part of the Hamiltonian, and explore the consequences at the Dirac point. It is found that the nonlocality predominantly influences the wave function rather than the energy spectrum: despite the persistence of the linear-in-momentum dispersion with a renormalized velocity, the wave functions no longer mirror those of SLG. Instead, an additional contribution to the phase difference between the sublattice components of the spinor emerges. This gives rise to interesting effects in scattering which are demonstrated with a simple example.
\end{abstract}

\maketitle

Twisted bilayer graphene (TBG) has attracted considerable attention in recent years. It arises when two sheets of single-layer graphene (SLG) are stacked and rotated by an arbitrary angle, resulting in a large-scale moir{\'e} pattern. At small angles of rotation, TBG exhibits a linear-in-momentum spectrum like single-layer graphene, albeit with a reduced velocity \cite{lopes2007graphene,trambly2010localization}, which goes to zero with concomitant band flattening at specific angles termed magic angles \cite{bistritzer2011moire}. With kinetic energy suppressed, electron-electron interactions dominate and drive the system to exotic phases which have been experimentally observed \cite{cao2018correlated,cao2018unconventional,lu2019superconductors,
yankowitz2019tuning,
chen2020tunable,choi2019electronic,jaoui2022quantum,
oh2021evidence,saito2020independent,stepanov2020untying,liu2021tuning,
lin2022spin,nuckolls2020strongly,pierce2021unconventional}---see Refs.~\onlinecite{andrei2020graphene,balents2020superconductivity,
bhowmik2023emergent} for a review on recent developments. This interesting phenomenon extends beyond TBG: Similar effects have been identified in other moir{\'e} materials involving more than two graphene layers \cite{chen2019evidence,chen2021electrically,liu2020tunable,shen2020correlated,
he2021symmetry,burg2019correlated,park2022robust,burg2022emergence} or constructed from alternative two-dimensional materials instead of graphene \cite{mak2022semiconductor}. 

A low-energy description of TBG can be obtained by expanding near the Dirac points of the individual layers. In this approximation, the individual layers are described by Dirac Hamiltonians, which are then coupled by an interlayer hopping term. Due to the emergence of a large moir{\'e} superlattice, the interlayer hopping term is spatially varying and shares the periodicity of the superlattice instead of the individual layers and has the form \cite{lopes2007graphene,bistritzer2011moire,mele2011band}
\begin{equation}
H_{\perp}(\mb{r})=\sum_{n=0}^2e^{i \delta\mathbf{K}_n\cdot\mathbf{r}}
T^n,
\label{interham}
\end{equation}
with  
\beq
T^n=\gamma\begin{pmatrix}
a & e^{-i 2\pi n/3}\\
 e^{i 2\pi n/3}& a 
\end{pmatrix},
\label{tn}
\eeq
where the matrix is written in the sublattice basis and the angle of rotation $\theta$ is taken to be zero when the two layers are in AA configuration. Here, $\mathcal{\gamma}$  quantifies the interlayer hopping strength (appropriately defined later), $\delta \mb{K}_n$ is the vector $\delta \mb{K}=\mathbf{K}^{\theta}-\mathbf{K}$ rotated by $2n\pi/3$ with $n=0,1,2$, $\mb{K}$ and $\mb{K}^\theta$ are the Dirac points of the two layers, respectively, and $a$ is a real number that captures the asymmetry in the AA- and AB-hoppings \cite{mele2011band}.

\begin{figure}
\centering
\includegraphics[width=0.49\textwidth]{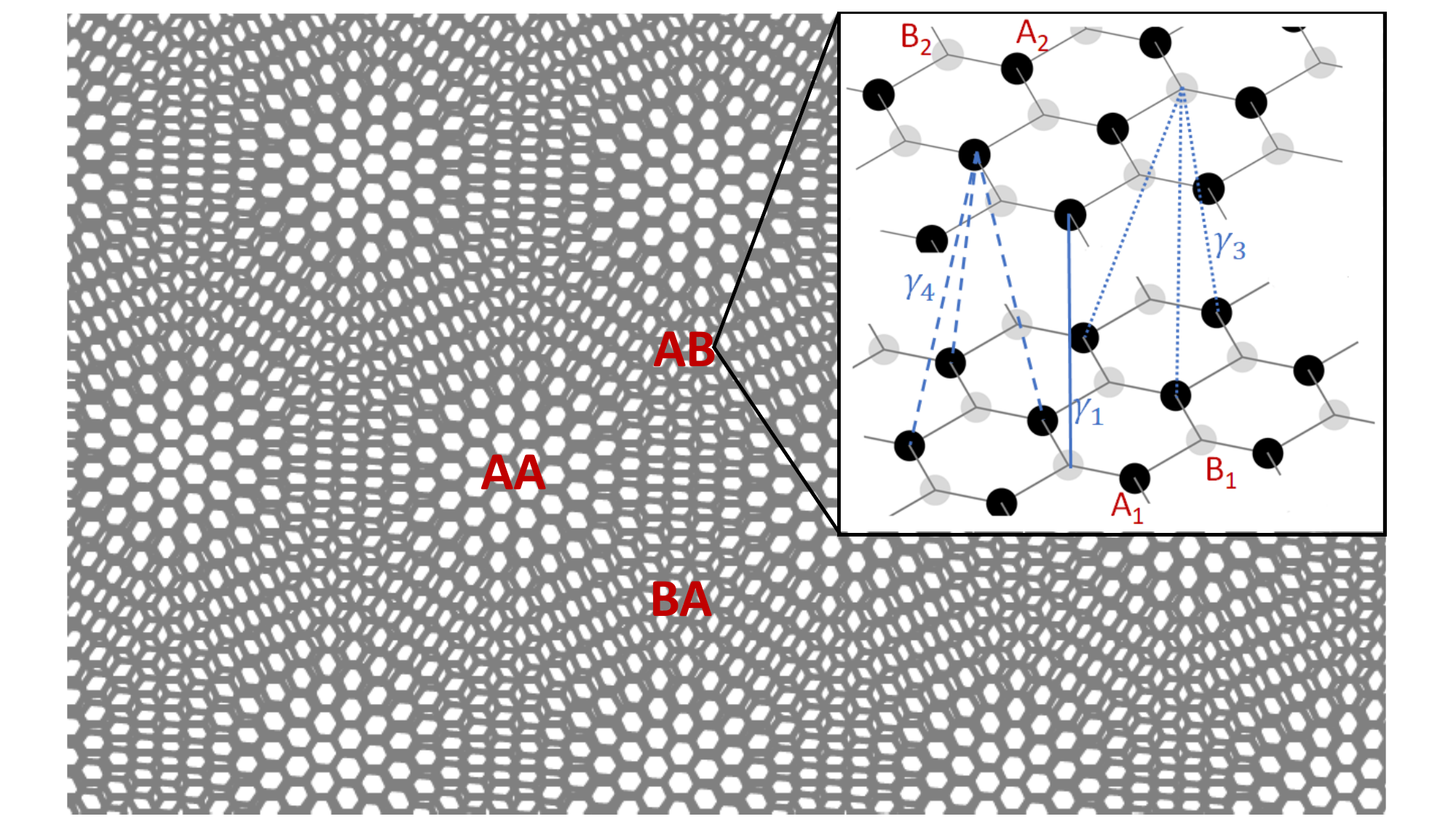}
\caption{Moir{\'e} pattern in TBG showing locally AA, AB and BA regions. The inset magnifies the AB region showing the atomic registry: In addition to the nn interlayer hopping parameter $\gamma_1$, there exists two nnn hopping parameters $\gamma_3$ and $\gamma_4$.}
\label{moire_ab}
\end{figure}

The interlayer Hamiltonian in (\ref{interham}) has a simple interpretation. In TBG, the local registry of atoms between the two layers changes smoothly in real space, with a periodicity of the moir{\'e} superlattice. As shown in Fig.~\ref{moire_ab}, it is locally AA-like in certain regions and AB(BA)-like in other regions. From geometry, the AA-like regions occur at $\mb{r}=0$ (and superlattice translations), whereas the AB-like regions arise at $\mb{r}= (4\pi/3\sqrt{3} \delta K^2) {\rm R}(\pi/6) \mathbf{\delta K}$, where $R(\varphi)$ denotes a rotation by angle $\varphi$ (and superlattice translations). At these $\mb{r}$ values, (\ref{interham}) reduces to
\begin{equation}
H_\perp(AA)=3a\gamma
\begin{pmatrix}
1& 0\\
0&1
\end{pmatrix},\ \ 
H_\perp(AB)=3\gamma 
\begin{pmatrix}
0& 0\\
1&0
\end{pmatrix}.
\label{aaab}
\end{equation}
These are exactly the forms of the interlayer hopping term expected in pristine AA and AB bilayer graphene \cite{mccann2013electronic}. Thus, (\ref{interham}) essentially interpolates between locally AA and AB(BA) regions, including the intervening regions where the local registry of atoms is a mix of both configurations.

While the above discussion lends a simple physical picture, it also points to an obvious omission. The expressions in (\ref{aaab}) are expected when one considers only the nearest-neighbor (nn) interlayer hopping. If the next-nearest-neighbor (nnn) interlayer hopping is included, the expressions change both qualitatively and quantitatively. For example, it is well known that the interlayer Hamiltonian in pristine AB bilayer graphene has the form \cite{mccann2013electronic}
\begin{equation}
H_\perp(AB)=
\begin{pmatrix}
-v_4 k_-& v_3 k_+\\
\gamma_1&-v_4 k_-
\end{pmatrix},
\label{abham}
\end{equation}
where $k_\pm=k_x\pm i k_y$, $\gamma_{1}$ is the nn hopping parameter and $v_{3,4}$ are two constants derived from the two nnn hopping parameters $\gamma_{3,4}$ shown in the inset to Fig.~\ref{moire_ab}. The inclusion of the nnn parameters is important for a variety of reasons \cite{mccann2013electronic}. First, the nnn parameters are not necessarily smaller than the nn parameters to warrant an omission: $\gamma_3$ is similar to $\gamma_1$ in magnitude. Second, the nnn parameters qualitatively modify the form of the interlayer hopping Hamiltonian---they make it $k-$dependent, and hence, nonlocal. And, third, they give rise to qualitatively new features at low energies, e.g., Lifshitz transitions. That (\ref{interham}) leads to (\ref{abham}), instead of (\ref{aaab}), at locally AB-like regions clearly indicates that \ref{interham}) is local and captures the effects of only nn parameters in TBG. It is, therefore, natural to ask what happens to TBG when nnn hopping parameters are included and the interlayer hopping terms become nonlocal.

A few recent studies have addressed the issue of nonlocal interlayer hopping using a variety of approaches \cite{guinea2019continuum,carr2019exact,garcia2021full,xie2021weak,kang2023pseudomagnetic}. These investigations converge on the common observation that such terms give rise to particle-hole asymmetry in the TBG spectrum. Notably, the effect manifests predominantly at the mini-Brillouin zone-$\Gamma$ point, leaving the Dirac point largely unaffected.  Consequently, it might seem that nonlocal interlayer hopping has no effect there. Here, I show that, while the energy spectrum does not get affected appreciably at the Dirac point, the wave function is modified significantly: it is no longer the same as that of SLG. More specifically, the phase difference between the sublattice components of the wave function acquires an extra contribution, which can lead to unusual consequences.

To include nonlocal terms in the interlayer Hamiltonian, one can adopt the same approach used in deriving the local version (\ref{interham}). Within the tight-binding scheme, a general expression of the interlayer hopping term can be written as
$H_\perp=\sum_{i,j,\alpha,\beta}t_\perp\left( \mr_{i\alpha}-\mr ^\theta_{j\beta}\right)\left[c_{\alpha}^\dagger( \mr_i)c_\beta(\mr^\theta_j)+\mrm{h.c.} \right]$.
Here, $c_{\alpha}^\dagger( \mr_i)$ creates an electron on one layer (unrotated) at lattice site $i$ and sublattice $\alpha$, $c_\beta(\mr^\theta_j)$ destroys an electron on the other layer (rotated) at lattice site $j$ and sublattice $\beta$, and $t_\perp\left( \mr_{i\alpha}-\mr^\theta_{j\beta}\right)$ denotes the spatially varying interlayer hopping parameter. For low-energy theory, one expands near the Dirac points of the two layers, $\mb{K}$ and $\mb{K}^\theta$, respectively. Focusing on one valley (say $\mb{K}$), define $c_{\alpha}( \mr_i)=e^{i\mb{K}\cdot\mr_i}\psi_\alpha(\mr_i)$, with $\psi_\alpha(\mr_i)=\frac{1}{\sqrt{\mathcal{V}}}\sum_{\mk}e^{i\mk\cdot\mr_i}\phi_\alpha(\mk)$, where $\mathcal{V}$ is the volume. This gives
\begin{eqnarray}
H_\perp&=&\frac{1}{\mathcal{V}}\sum_{i,j,\alpha,\beta,\mk,\mk ^\theta}t_\perp\left( \mr_{i\alpha}-\mr ^\theta_{j\beta}\right)e^{-i(\mb{K}+\mk)\cdot\mr_i}e^{i(\mb{K}^\theta+\mk ^\theta)\cdot\mr ^\theta_j}\nonumber\\
&&\phi_\alpha^\dagger(\mk)\phi_\beta(\mk ^\theta)+\mrm{h.c.}
\label{inter2}
\end{eqnarray}
Assuming translational invariance, $t_\perp\left( \mr_{i\alpha}-\mr ^\theta_{j\beta}\right)$ can be expressed in terms of its Fourier transform:   
\beq
t_\perp\left( \mr_{i\alpha}-\mr ^\theta_{j\beta}\right)
=\frac{1}{\mathcal{V}}\sum_{\mb{q}}\tilde{t}^{\alpha\beta}_\perp(\mb{q})e^{i\mb{q}\cdot(\mr_{i}-\mr ^\theta_{j}+\mb{\ell}_\alpha-\mb{\ell} ^\theta_\beta)},
\eeq
where $\mb{\ell}_1=\mb{0}$ and $\mb{\ell}_2=\mb{\ell}$ are the vectors denoting the position of the basis atoms in a single layer. Using this in (\ref{inter2}) and summing---first over $\mr_i$ and $\mr ^\theta_j$ and subsequently over $\mb{q}$---yields
\begin{eqnarray}
H_\perp &=&\sum_{\substack{\alpha,\beta\\\mk,\mk ^\theta\\\mb{G},\mb{G}^\theta}}\tilde t^{\alpha\beta}_\perp(\mb{K} +\mb{G}+\mk)e^{i(\mb{K} +\mb{G}+\mk)\cdot\mb{\ell}_\alpha}e^{-i(\mb{K}^\theta+\mb{G}^\theta+\mk ^\theta)\cdot\mb{\ell}^\theta_\beta}\nonumber\\
&\times &\delta_{\mk+\mb{K}+\mb{G},\mk ^\theta+\mb{K}^\theta+\mb{G}^\theta}\phi_\alpha^\dagger(\mk)\phi_\beta(\mk ^\theta)+\mrm{h.c.},
\label{inter3}
\end{eqnarray}
where $\mb{G}$ and $\mb{G}^\theta$ are the reciprocal lattice vectors of the unrotated and rotated layers, respectively. This is as far as one can go on general grounds. A further simplification occurs, however, if one makes the following two approximations:
(i) Restrict the summation over $\mb{G}$ and $\mb{G}^\theta$ to only those values which result in $|\mb{K}+\mb{G}|=|\mb{K}|$: This is justified since, in reality, $\tilde{t}(\mb{q})$ reduces rapidly with increasing $|\mb{q}|$ and, therefore, only those terms in the summation that come with the Fourier coefficient $\tilde{t}^{\alpha\beta}(|\mb{K}|)$ need to be considered. Define  $\tilde{t}^{\alpha\beta}(|\mb{K}|)\equiv \gamma$ when $\alpha\ne\beta$ and $\equiv a\gamma$ when $\alpha=\beta$.
(ii) Neglect $k$ and $k^\theta$ in the first line: This is justified because low-energy approximation implies $|\mk|\ll|\mb{K}+\mb{G}|$ and $|\mk^\theta|\ll|\mb{K}^\theta+\mb{G}^\theta|$.
Using these two approximations, (\ref{inter3}) reduces to
\beq
H_\perp=\sum_{\substack{\alpha,\beta\\\mk,\mk^\theta}}\sum_nT^n_{\alpha\beta}\delta_{\mk,\mk ^\theta+\delta\mb{K}_n}\phi_\alpha^\dagger(\mk)\phi_\beta(\mk ^\theta)+\mrm{h.c.},
\label{interhamk}
\eeq
The interlayer hopping term written at the outset in (\ref{interham}) is simply the Fourier transform of (\ref{interhamk}) to real space.

It is clear that approximation (ii) needs to be discarded to include nonlocality in interlayer hopping. To that effect, one can expand the terms in the first line in (\ref{inter3}) to first order in $\mk,\mk^\theta$. Then,
\begin{eqnarray}
\tilde t^{\alpha\beta}_\perp(\mb{K} +\mb{G}+\mk)&\approx &  \tilde t^{\alpha\beta}_\perp(\mb{K} +\mb{G})+\mk\cdot \partial_\mk \tilde{t}^{\alpha\beta}_\perp(\mk)|_{\mk =\mb{K} +\mb{G}}\nonumber\\
&=&
    \begin{cases}
    \gamma+\tilde v\mk\cdot\frac{(\mb{K}+\mb{G})}{|\mb{K}+\mb{G}|}& \alpha\ne\beta\\
   
      a\gamma+b\tilde v\mk\cdot\frac{(\mb{K}+\mb{G})}{|\mb{K}+\mb{G}|}& \alpha=\beta,
    \end{cases} 
\label{texpan}
\end{eqnarray}
where it has been assumed $\tilde{t}_\perp (\mb{q})=\tilde{t}_\perp(|\mb{q}|)$ and $a$ and $b$ are real numbers. One should also expand the exponential terms, but they give rise to terms that are smaller by a factor $l/L \ll 1$, where $l$ and $L$ are the lattice constants of a single layer and moir{\'e} lattice, respectively, and can be ignored. Approximation (i) is still valid and can be used to choose only those $\mb{G}$ vectors that maintain $|\mb{K}+\mb{G}|=|\mb{K}|$. Thus, one arrives at (\ref{interhamk}) once again, except that the matrix $T^n$ is now replaced by
\begin{eqnarray}
T^n(\mk)&=&\gamma
\begin{pmatrix}
a & e^{-i 2\pi n/3}\\
 e^{i 2\pi n/3}& a 
\end{pmatrix}\nonumber\\
&+&
\tilde{v}\mk\cdot \mb{K}_n\begin{pmatrix}
b & e^{-i 2\pi n/3}\\
 e^{i 2\pi n/3}& b 
\end{pmatrix}.
\label{tnk}
\end{eqnarray}
The above expression is the nonlocal generalization of the local form in (\ref{tn}) and is similar to the one found in Ref.\cite{xie2021weak}. Here, $\gamma$ and $\tilde{v}$ parametrize the strength of the local and nonlocal hoppings, respectively, while $a$ and $b$ capture the asymmetry in the AA- and AB-hoppings, in the local and nonlocal contributions, respectively. It can be easily verified that using this revised form in (\ref{interham}) and evaluating it at the locally AB-like regions, one gets the expression in (\ref{abham}) instead of (\ref{aaab}), provided one makes the following identification:
\beq
3b\tilde v/2\rightarrow -v_4;\ \ 3\gamma\rightarrow\gamma _1; \ \ 3\tilde v/2\rightarrow v_3.
\label{ABparametersmap}
\eeq
The mapping in (\ref{ABparametersmap}), alongside (\ref{aaab}), establishes a connection between the parameters employed in (\ref{tnk}) and those used in pristine AA and AB bilayer graphene. The parameters $a$ and $b$ deserve particular attention: While $a$ distinguishes the nn hopping in AA and AB configurations, $b$ arises due to the difference between the two kinds of nnn hopping in AB bilayer graphene, $\gamma_3$ and $\gamma_4$ (see Fig.~\ref{moire_ab}). Therefore, $a$ and $b$ are expected to deviate from unity. Importantly, this occurs even in the absence of lattice relaxation, which is customarily attributed as the source of this deviation. Nonetheless, their precise values are significantly influenced by lattice relaxation in TBG which may lead to substantial deviations from what is expected in pristine AA and AB bilayer graphene.

\begin{figure}
\centering
\subfigure[]{\includegraphics[width=0.45\textwidth]{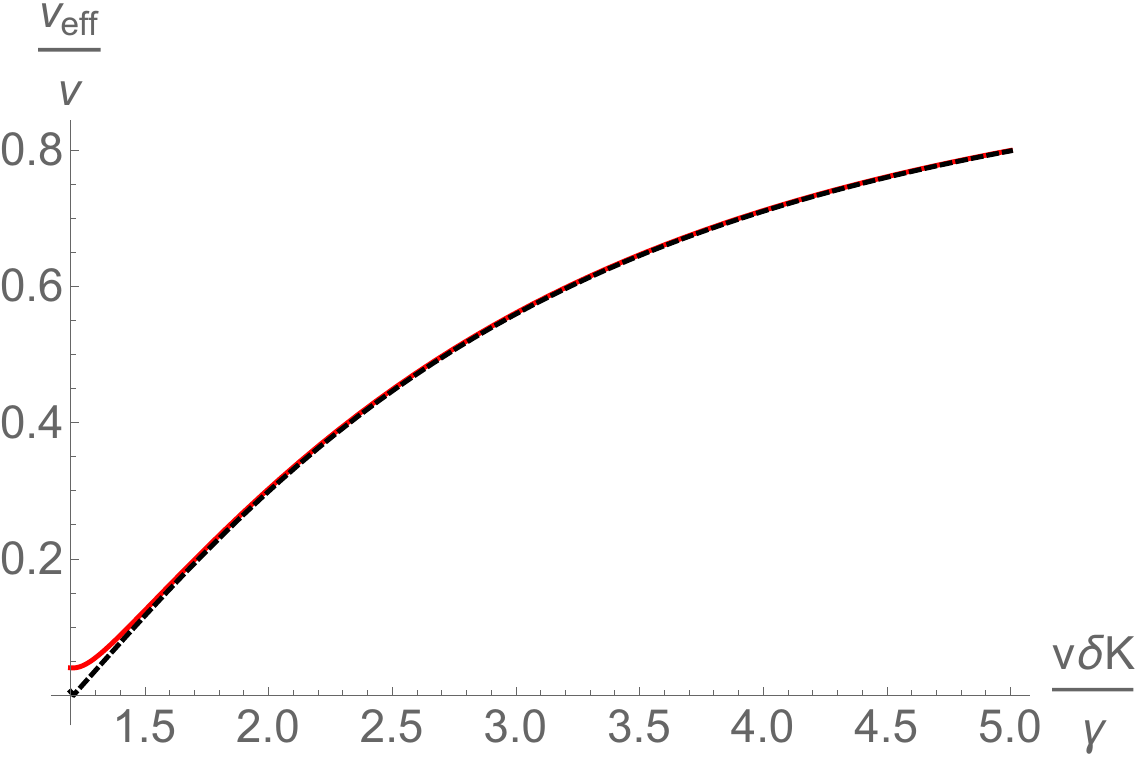}}
\quad
\subfigure[]{\includegraphics[width=0.45\textwidth]{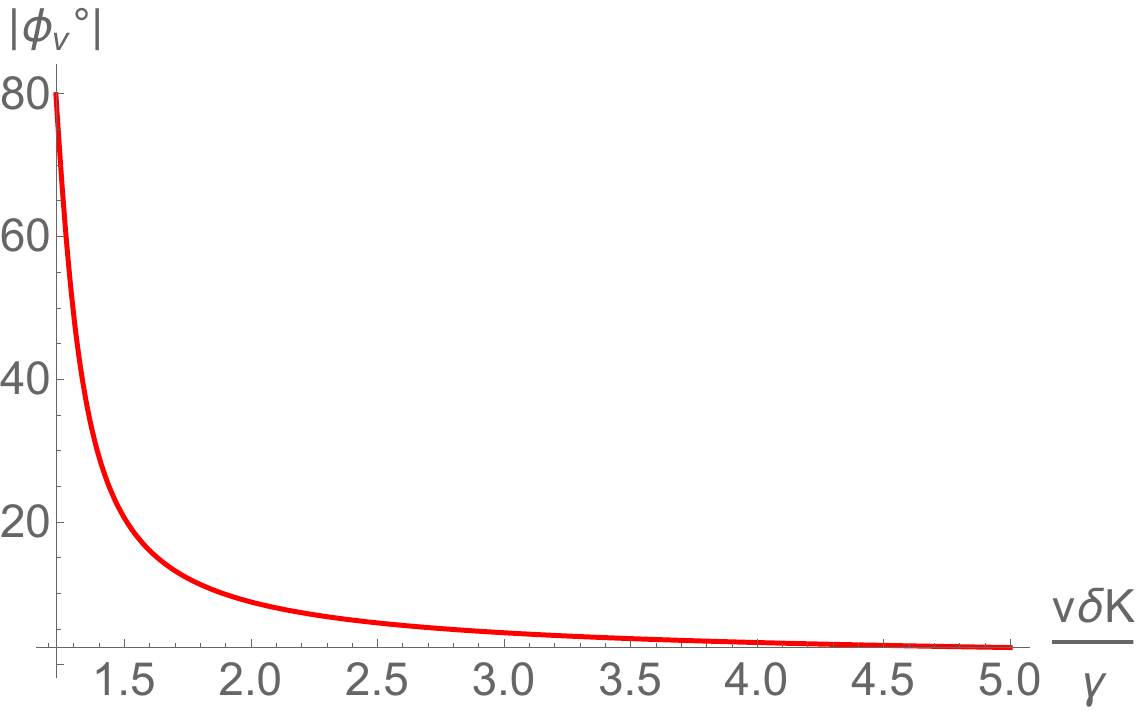}}
\caption{(a) Effective velocity $v_{\mrm{eff}}$ and (b) angle $\phi_v$ according to Eqs.~(\ref{veff}) and (\ref{phiv}), respectively, for $a=0.7$, $b=0.5$, and $\tilde{v}/v=0.1$. The dashed curve in (a) corresponds to $\tilde{v}=0$.}
\label{velphiv}
\end{figure}

Let us now explore the consequences of the nonlocal interlayer coupling in TBG, focusing on the low-energy physics near the Dirac point. Following Ref. \onlinecite{bistritzer2011moire}, in the limit $\gamma/v\delta K<1$, where $v$ is the single-layer Dirac velocity, the following truncated Hamiltonian describes TBG: 
\begin{eqnarray}
H_{\mb{k}}&=&
\begin{pmatrix}
h(\mb{k})&T_1(\mk)&T_2(\mk)&T_3(\mk)\\
T_1^{\dagger}(\mk)&h(\mb{k}+\mb{\delta K}_1)&0&0\\
T_2^{\dagger}(\mk)&0&h(\mb{k}+\mb{\delta K}_2)&0\\
T_3^{\dagger}(\mk)&0&0&h(\mb{k}+\mb{\delta K}_3)
\end{pmatrix},\nonumber\\
\label{hamtrunc}
\end{eqnarray}
where $h(\mb{q})$ is the single layer Dirac Hamiltonian with Dirac momentum $\mb{q}$ \footnote{The Dirac Hamiltonians $h(\mk+\delta\mb{K}_n)$ belonging to the rotated layer should strictly be replaced by rotated Dirac Hamiltonians: $e^{-i\sigma_z\theta/2}h(\mk+\delta\mb{K}_n)e^{i\sigma_z\theta/2}$. However, at small values of $\theta$, such changes have a negligible effect and can be ignored, as shown in Ref.~\cite{bistritzer2011moire}}. Carrying out a perturbative calculation to linear order in $\mk$, as outlined in Ref. \onlinecite{bistritzer2011moire}, the Hamiltonian (\ref{hamtrunc}) reduces to the following effective Hamiltonian: 
\beq
H_\mk=\begin{pmatrix}
0& v_-k_-\\
v_+k_+&0
\end{pmatrix},
\label{hameffk}
\eeq
where
\beq
\frac{v_\pm}{v} =\frac{1-\frac{3a^2\gamma^2}{ v^2\delta K^2}\mp i(ab-1)\frac{3\gamma}{v\delta K}\frac{\tilde{v}}{v}}{1+\frac{3(a^2+1)\gamma^2}{v^2\delta K^2}}\label{vpm}.
\eeq
It is observed that the main effect of the nonlocal coupling is to make $v_\pm$ complex which are otherwise real and equal to each other in the local scenario (when $\tilde{v}=0$) \cite{bistritzer2011moire}. Interestingly, for $\mrm{Im}[v_+]$ to be nonzero, in addition to $\tilde{v}\ne 0$, one also requires $ab\ne 1$, implying that for the effects of nonlocal coupling to appear, it is necessary to have an asymmetry in the interlayer hopping amplitude in the AA- and AB-like regions. The  eigenvalues and eigenvectors corresponding to (\ref{hameffk}) are:
\begin{eqnarray}
E_\pm &=&\pm v_{\mrm{eff}}k,\\
\psi_\pm &=&\frac{1}{\sqrt{2}}
\begin{pmatrix}
1\\
\pm e^{i(\phi_\mk+{\phi_v})}
\end{pmatrix},
\end{eqnarray}
where 
\begin{eqnarray}
v_{\mrm{eff}}&=&\sqrt{\mrm{Re}[v_+]^2+\mrm{Im}[v_+]^2},\label{veff}\\
\phi_v&=&\mrm{tan}^{-1}\left(\frac{\mrm{Im}[v_+]}{\mrm{Re}[v_+]}\right),\label{phiv}
\end{eqnarray}
and $\phi_\mk=\mrm{tan}^{-1}\left(\frac{k_y}{k_x}\right)$.
The energy spectrum remains linear in momentum like SLG, but the eigenfunction is manifestly different from that of a single layer, thanks to the appearance of the extra phase $\phi_v$ in the second component of the spinor. Indeed, the eigenfunction of TBG is a linear superposition of the single-layer eigenfunctions $\psi_{0\pm}$: $\psi_\pm=\left(\frac{1\pm e^{i\phi_v}}{2}\right)\psi_{0+}+\left(\frac{1\mp e^{i\phi_v}}{2}\right) \psi_{0-}$. This is in striking contrast to the local case ($\phi_v=0$) where the TBG wave functions are identical to that of SLG.

How large is the effect of nonlocal coupling? Using $a\approx 0.7$, $b\approx 0.5$ \cite{kang2023pseudomagnetic,moon2012energy,fang2016electronic} and $\tilde{v}/v\approx 0.1$\cite{mccann2013electronic} in (\ref{vpm}),  $\mrm{Re}[v_+/v]=\left(1-\frac{1.47\gamma^2}{ v^2\delta K^2}\right)\big/\left(1+\frac{4.47\gamma^2}{ v^2\delta K^2}\right)$ and $\mrm{Im}[v_+]= \left(0.195\frac{\gamma}{v\delta K}\right)\big/\left(1+\frac{4.47\gamma^2}{ v^2\delta K^2}\right)$. At all values of $\frac{\gamma}{v\delta K}\ll 1$, $\mrm{Im}[v_+]\ll\mrm{Re}[v_+]$ and no additional effect appears due to nonlocal coupling. However, as the rotation angle decreases ($\frac{\gamma}{v\delta K}$ increases) and one approaches the vicinity of the first magic angle, $\mrm{Re}[v_+]$ becomes small, approaching zero allowing $\mrm{Im}[v_+]$ to contribute. Note that the latter is still numerically small, and does not significantly modify $v_{\mrm{eff}}$. This is shown in Fig.~\ref{velphiv}(a). However, its effect on $\phi_v$ is substantial: no matter how small $\mrm{Im}[v_+]$ is, because $\mrm{Re}[v_+]$ is approaching the value zero with decreasing angle of rotation, $\phi_v$ approaches the value $90^\circ$, as shown in Fig.~\ref{velphiv}(b). Thus, in essence,  the nonlocal part, through $\mrm{Im}[v_+]$, introduces $\phi_v$, while the local part, via $\mrm{Re}[v_+]$, dictates its magnitude. This duality ensures the robustness of the effect, rendering it qualitatively immune to the parameters used and the approximation employed in (\ref{hamtrunc}). Indeed, approaching the magic angle, the truncated Hamiltonian (\ref{hamtrunc}), correct only up to $O(\gamma^2/v^2\delta K^2)$, becomes inadequate and terms of order $O(\gamma^4/v^4\delta K^4)$ and higher need to be included. However, such terms only modify the angle at which $\mrm{Re}[v_+]$ goes to zero but does not prevent it from doing so \cite{bistritzer2011moire}, implying that a nonzero $\phi_v$, increasing with decreasing angle, is intrinsic to TBG.

\begin{figure}
\centering
\subfigure[]{\includegraphics[width=0.49\textwidth]{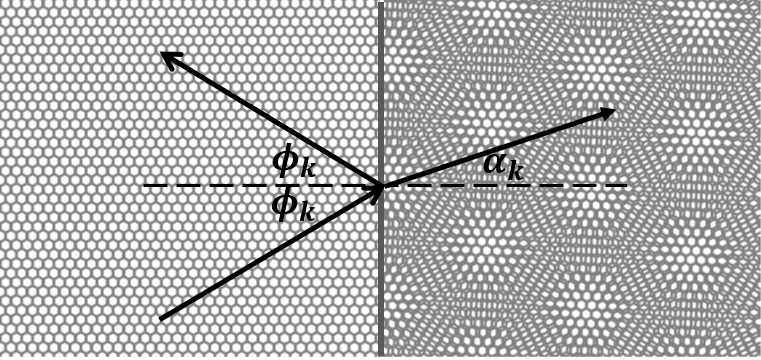}}
\quad
\subfigure[]{\includegraphics[width=0.5\textwidth]{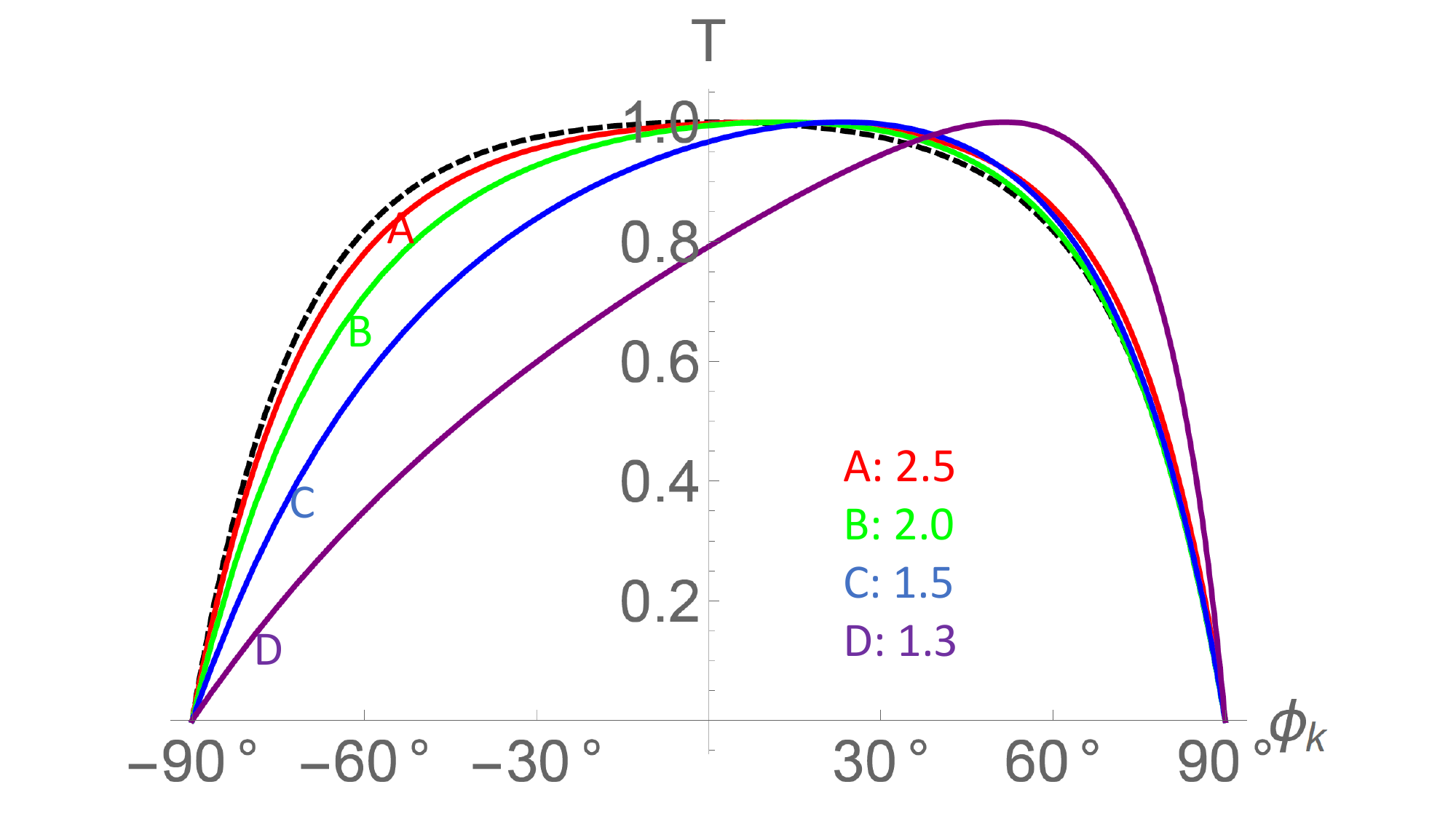}}
\caption{(a) A SLG-TBG heterostructure where an electron wave incident at the interface from the SLG side, gets partly reflected back to SLG and partly transmitted to TBG. (b) The transmission coefficient according to Eq.~(\ref{trans}) as a function of the incident angle. The parametrs $a$, $b$, and $\tilde v/v$ are the same as used in Fig.~\ref{velphiv}. The different curves represent different angles of rotation of TBG represented by $v\delta K/\gamma$ whose values are given in the legend. The dashed curve corresponds to $\tilde v=0$ and $v\delta K/\gamma=2.5$.}
\label{interface}
\end{figure}

The modification of the wave function is expected to lead to qualitatively new features, particularly in scattering. I demonstrate this with a simple example. Consider a heterostructure of SLG and TBG as shown in Fig.\ref{interface}(a) with the interface at $x=x_0$. For simplicity, I assume the interface to be sharp. Electrons incident from the SLG side will scatter at the interface resulting in a part of it being reflected back to SLG with reflection amplitude $r$ and the rest transmitted to TBG with transmission amplitude $t$. For an incident electron wave at an angle $\phi_\mk$ with the normal to the interface, the continuity of the wave function at the interface leads to the following equations:
\begin{eqnarray}
e^{ik_xx_0}+r e^{-ik_xx_0}&=&te^{iq_xx_0},\\
e^{ik_xx_0+i\phi_{\mk}}-r e^{-ik_xx_0-i\phi_{\mk}}&=&te^{iq_xx_0+i\alpha_\mk+i\phi_v},
\end{eqnarray}
where $\alpha_\mk=\mrm{sin}^{-1}(\frac{v_{\mrm{eff}}}{v}\mrm{sin}\phi_\mk)$ and $q_x=k\sqrt{(\frac{v_{\mrm{eff}}}{v})^2-\mrm{sin}^2\phi_\mk}$, with $v_{\mrm{eff}}$ given by (\ref{veff}). The reflection and transmission probabilities, $R$ and $T$, respectively, are obtained by solving the above equations:
\begin{eqnarray}
R&=&|r|^2=\Bigg\lvert\frac{1-e^{i\alpha_\mk+i\phi_v-i\phi_\mk}}{1+e^{i\alpha_\mk+i\phi_v+i\phi_\mk}}\Bigg\rvert^2, \\
T&=&1-R.
\label{trans}
\end{eqnarray}
The unusual nature of the scattering at the interface becomes evident when one considers the angle of incidence at which there is perfect transmission. In the local approximation, when $\phi_v=0$, it is seen that $T=1$ when $\phi_\mk=0$, i.e., electrons incident normally on the interface do not suffer any reflection. This is expected because, in the local approximation, a TBG is effectively a SLG. This no longer holds in the nonlocal scenario. With a nonzero $\phi_v$, electrons incident normally no longer is transmitted entirely, with a part of them getting reflected as shown in Fig.~\ref{interface}(b). Remarkably, transmission becomes perfect at an oblique incidence, at an angle $\phi_\mk^\ast$ given by
\beq
 \phi_\mk^\ast=\mrm{sin}^{-1}\left[\left\{1+\left(\frac{\mrm{cos}\phi_v-v_{\mrm{eff}}/v}{\mrm{sin}\phi_v}\right)^2\right\}^{-1/2}\right],
\eeq
whereas, at normal incidence, transmission is given by
\beq
T_{\phi_\mk=0}=1-\mrm{tan}^2(\phi_v/2),
\eeq
which goes to zero as one approaches the magic angle (i.e., $\phi_v$ approaches $90^\circ$). This behavior arises purely from quantum interference between the wave functions on either side of the heterostructure.

The findings open several possibilities for future studies. As in TBG, nonlocal interlayer hopping is anticipated in other moir{\'e} materials and it remains to be seen how the low-energy physics is affected in those systems.  Additionally, all scattering matrix elements, in general, will be modified due to the nonlocal term, which is expected to lead to new features in transport and interaction-driven physics. 

In summary, I have explored the effects of nonlocal interlayer hopping in TBG. It is found that at the Dirac point, it has a negligible effect on the energy spectrum but has a significant impact on the wave function. The wave function is no longer identical to that of SLG; specifically, the phase difference between the sublattice components of the spinor wave function acquires an extra contribution that is dependent on both the strength of the nonlocal hopping as well as the sublattice asymmetry inherent in TBG.

\begin{acknowledgments}
I would like to thank DST SERB, India for financial support via Grant No. CRG/2021/005453.
\end{acknowledgments}

\end{document}